\documentclass{article}
\usepackage{amsfonts,amssymb, amsmath,mathrsfs}
\usepackage[english]{babel}

\def\g{{\gamma}}

\def\bq{ \begin{equation} }
\def\eq{ \end{equation} }
\def\ben{ \begin{eqnarray} }
\def\en{ \end{eqnarray} }

\newtheorem{prop}{Proposition}
\begin{document}
%%%%%%%%%%%% TITLE %%%%%%%%%%%%%%

\title{On generalized nonholonomic Chaplygin sphere problem}
\author{A.V. Tsiganov \\
\it\small
St.Petersburg State University, St.Petersburg, Russia\\
\it\small e--mail: andrey.tsiganov@gmail.com}
\date{}
\maketitle

\begin{abstract}
We discuss  linear in momenta Poisson structure for the generalized nonholonomic
Chaplygin sphere problem and prove that it is non-trivial deformation of the canonical
Poisson structure  on $e^*(3)$.
 \end{abstract}

%%%%%%%%%%%%%%%%%%%%%%%%%%%%%%%%%%%%%%%%%
\section{Introduction}
\setcounter{equation}{0}
Let us  consider a rolling of dynamically asymmetric and balanced spherical rigid body, the so-called Chap\-ly\-gin ball, over an
 absolutely rough fixed  sphere with  radius $a$  \cite{bf95}.   At $a\to \infty$  one gets  a Chaplygin
problem on a non-homogeneous sphere rolling over a horizontal plane without slipping  \cite{ch03}.

Since  slipping at
 the contact points is absent, its  velocity vanishes and we have the following nonholonomic constraint
\bq\label{rel-ch}
v+\omega\times r=0\,.
\eq
Here  $\omega$ and $v$ are the angular velocity and velocity of the center of mass of the ball,
$r$ is the vector joining the center of mass with the contact point and  $\times$ means the vector product in $\mathbb R^3$. Mass,  inertia tensor and radius of the rolling ball  will be denoted by  $m$, $\mathbf I = \mathrm{diag}(I_1, I_2, I_3 )$ and $b$, respectively.

 According to  \cite{bf95}, the angular momentum  $M$ of the ball with respect to the contact point with the sphere  is equal to
  \bq\label{om-m}
M=(\mathbf I+d\mathbf E)\,\omega-d(\g,\omega)\g\,,\qquad \qquad d=mb^2.
\eq
Here $\g$ is the unit normal vector to the fixed sphere at the contact point, $\mathbf E$ is the unit matrix and $(.,.)$ means the standard scalar product in $\mathbb R^3$. All the vectors are expressed in the so-called body frame,  which is firmly attached to the ball,
its origin is located at the center of mass of the body, and its axes coincide with the principal inertia axes of the body.

After elimination of the Lagrangian multiplier according to  \cite{bf95}, one gets the following  reduced equations of motion
\bq\label{m-eq}\dot M=M\times \omega\,,\qquad \dot \g=\kappa \g\times \omega\,,\qquad\mbox{where}\qquad
\kappa=\dfrac{a}{a+b}.
\eq
These equations  possess three integrals of motion
\bq\label{3-int}
H_1=(M,\omega)\,,\qquad H_2=(M,M)\,,\qquad  C_1=(\g,\g)\,,
\eq
and invariant measure
\bq\label{rho}
\quad \mu= {\mathrm g^{-1}(\g)\,}\,\mathrm d\g \,\mathrm dM\,,\qquad
\mathrm g(\g)=\sqrt{1-d ( \g,\mathbf A \g)}\,,
\eq
where
\[
\mathbf A=\left(
            \begin{array}{ccc}
              a_1 & 0 & 0 \\
              0 & a_2 & 0 \\
              0 & 0 & a_3
            \end{array}
          \right)=(\mathbf I+d\mathbf E)^{-1}=\left(
            \begin{array}{ccc}
              \frac{1}{I_1+d} & 0 & 0 \\
              0 & \frac{1}{I_2+d} & 0 \\
              0 & 0 & \frac{1}{I_3+d}
            \end{array}
          \right)\,,
\]
At  $\kappa=\pm 1$  one more integral of motion  exists
\bq\label{4-int}
C_2=(\g,\mathbf B M)\,,\qquad
\mathbf B=\left(
            \begin{array}{ccc}
              b_1 & 0 & 0 \\
              0 & b_2 & 0 \\
              0 & 0 & b_3
            \end{array}
          \right)=\mbox{tr}\, \mathbf A^{-1}+(\kappa-1)\mathbf A^{-1}\,.
\eq
It is easy to see,  that  in the Chaplygin case $a\to \infty$ we have $\kappa=1$.

At $\kappa=-1$ we have generalised Chaplygin sphere problem or so-called Borisov-Mamaev-Fedorov system, see \cite{bf95} and \cite{bm08b}.

\section{The Poisson brackets}
\setcounter{equation}{0}
At $\kappa=\pm 1$ six equations of motion  (\ref{m-eq})  possess four integrals of motion and an invariant measure.
Thus, by the Euler-Jacobi theorem, they are integrable in quadratures. It allows us to suppose that
 common level surfaces of integrals form  a direct sum of  symplectic and lagrangian foliations of
 dual dynamical system which is hamiltonian with respect to the Poisson bivector $P$, so  that
\bq\label{geom-eq}
[P,P]=0\,,\qquad P\mathrm dC_{1,2}=0\,,\qquad (P\mathrm dH_1,\mathrm dH_2)\equiv\{H_1,H_2\}=0\,.
\eq
Here  $[.,.]$ is the Schouten  bracket. In fact, we suppose that the Euler-Jacobi  integrability
 of non-Hamiltonian system (\ref{m-eq}) is equivalent to the Liouville integrability of the
 dual Hamiltonian dynamical system with the same integrals of motion, see \cite{ts11}.

The first equation in (\ref{geom-eq}) guaranties that $P$ is a Poisson bivector. In the second equation  we define  two Casimir elements $C_{1,2}$  of $P$ and assume that rank$P=4$. It is a necessary condition because by fixing its values one gets  four dimensional symplectic phase space of the desired Hamiltonian system. The third equation provides that integrals $H_{1,2}$ are in  involution with respect to the Poisson
bracket associated with $P$ and, therefore,  that they  form a lagrangian foliation.

In order to compare Poisson structures at $\kappa=\pm1$ we briefly remind some known facts about
  linear in momenta  $M$ solutions $P$ of  (\ref{geom-eq}) associated with the Chaplygin sphere problem at $\kappa=1$ following to \cite{bm01,bm08b,ts12rd,ts12}.

\subsection{Chaplygin sphere, $\kappa=1$\,.}
According to  \cite{bm01},  integrals of motion  (\ref{3-int}-\ref{4-int}) are in  involution  with respect to the following Poisson brackets
\bq\label{3-br}\\
\{M_i,M_j\}_g=\varepsilon_{ijk}\left(\mathrm g{M_k}-\dfrac{d(M, \mathbf A \g)}{\mathrm g} \g_k\right),\quad
 \{M_i, \g_j\}_g=\varepsilon_{ijk}\mathrm g \g_k\,,\quad
\{ \g_i, \g_j\}_g=0,
\eq
where $\varepsilon_{ijk}$ is a totally skew-symmetric tensor. These brackets have the necessary Casimir functions $C_{1,2}$.

In variables  $x=(\g_1,\g_2,\g_3,M_1,M_2,M_3)$ initial equations of motion  (\ref{m-eq})  have the form
 \begin{equation}\label{new-eq}
\dfrac{\mathrm dx_k}{\mathrm dt}\equiv X_k=\mathrm g^{-1}\{H,x_k\}_g\,,\qquad\mbox{where}\qquad H=\dfrac{H_1}{2}\,.
\end{equation}
After a change of time
 \begin{equation} \label{time-ch}
\mathrm dt\to\mathrm  g\mathrm dt
  \end{equation}
these equations becomes Hamiltonian equations with respect to the Poisson brackets  (\ref{3-br}).
It means that  initial non-Hamiltonian vector field $X$ is the conformally Hamiltonian vector field
\[X =\mathrm g^{-1}(x)\,\hat{X}\,,\qquad\mbox{where }\qquad \hat{X}=P_g\,\mathrm dH\,.\]

The Poisson brackets  (\ref{3-br}) can be easily obtained via trivial deformations of the canonical Poisson brackets and the standard momentum map theory. Namely, let $Q$ be a $n$-dimensional smooth manifold. Its cotangent bundle $T^*Q $ is naturally endowed with the Liouville $1$-form $\theta$ and symplectic $2$-form $\Omega= d\theta$, whose associated Poisson bivector will be denoted with $P$.
In local symplectic coordinates on $T^\ast Q$
\[z=(q,p)=(q_1,\dots,q_n,p_1,\dots,p_n)\]
they have the following local expressions
 \[\theta=p_1\mathrm dq_1+\ldots p_n\mathrm dq_n\,,\qquad \Omega=\mathrm d\theta
 =p_1\wedge q_1+\cdots p_n\wedge q_n.\]
Let us substitute the scaling momenta
\bq\label{p-trans}
p_k\to g(q)\,p_k\,\qquad k=1,\ldots,n,
\eq
into the Liouville and symplectic forms
 \[ \theta_g=g(q)\Bigl(p_1\mathrm dq_1+\ldots p_n\mathrm dq_n\Bigr)\,,\qquad \Omega\to \Omega_g=\mathrm d\theta_g\,.\]
 The corresponding Poisson bracket
\bq\label{g-br}
\{q_i,q_j\}_g = 0\,,\qquad \{q_i,p_j\}_g=g\delta_{ij}\,,\quad
\{p_i,p_j\}_g=\partial_j g\,p_i-\partial_i g\,p_j\,,
\eq
is a trivial deformation of canonical Poisson bracket
\[
\{q_i,q_j\} = 0\,,\qquad \{q_i,p_j\}=\delta_{ij}\,,\quad
\{p_i,p_j\}=0\,
\]
in the Poisson-Lichnerowicz cohomology \cite{Turiel}. Here $ \partial_k=\partial/\partial q_k$.

Now let us  identify $Q$ with a two dimensional sphere  $S^2$ embedded into  $\mathbb R^3$,  so that $q_i=\g_i$, $i=1,2,3$.  The standard momentum map
\[
\phi:\qquad(p,\g)\in T^* S^2 \to (M,\g) \in e^*(3)=so(3)\ltimes \mathbb R^3\,,\qquad
\]
defined by the vector product
\bq\label{l-en}
M=\g\times p\,,
\eq
maps our  trivial deformation (\ref{g-br}) into the following Poisson brackets on the Lie algebra  $e^*(3)$
\ben\label{pg-e3b}
&&\bigl\{M_i\,,M_j\,\bigr\}_g=\varepsilon_{ijk}\,\left(g(\g)M_k
+\g_k\sum_{m=1}^3M_m\partial_m g(\g)\right)
\,,\\
&&
\bigl\{M_i\,,\g_j\,\bigr\}_g=\varepsilon_{ijk}\,g(\g)\,\g_k \,,
\qquad
\bigl\{\g_i\,,\g_j\,\bigr\}_g=0\,.
 \nonumber
\en
As above, here we use an abbreviation  $\partial_m=\partial/\partial \g_m$.
\begin{prop}
If we identify $g(\g)$ with  $\mathrm g(\g)$ (\ref{rho}), then
the Poisson brackets (\ref{pg-e3b}) coincide with  the Poisson brackets (\ref{3-br}).
\end{prop}
So, the Poisson brackets  (\ref{3-br}) are trivial deformations of canonical ones. Consequently,
according to  \cite{ts12}, change of variables
\begin{equation}\label{trans-lmf}
\begin{array}{l}
L_1=\mathrm g^{-1}\left(M_1-\dfrac{b\g_1}{(\g,\g)}\left(1+\dfrac{\g_3^2}{\nu}\right)\,
\right)+\dfrac{c\g_1}{\g_1^2+\g_2^2}\,,
\\
\\
L_2=\mathrm g^{-1}\left(M_2-\dfrac{b\g_2}{(\g,\g)}\left(1+\dfrac{\g_3^2}{\nu}\right)\,
\right)+\dfrac{c\g_2}{\g_1^2+\g_2^2}\,,
\\
\\
L_3=\mathrm g^{-1}\left(M_3
-\dfrac{b\g_3}{(\g,\g)}\left(1-\dfrac{\g_1^2+\g_2^2}{\nu}\right)\,
\right)\,,
\end{array}
 \end{equation}
where  \[ b=(\g,M)\,,\qquad c= (\g,L)\, \qquad\mbox{and}\qquad \nu=\g_1^2+\g_2^2-d(\g,\g)(a_1\g_1^2+a_2\g_2^2)\,,\]
allows us to reduce the deformed Poisson brackets (\ref{3-br}) to the canonical Lie-Poisson brackets on the
Lie algebra $e^*(3)$
  \begin{equation}\label{e3}
\bigl\{L_i\,,L_j\,\bigr\}=\varepsilon_{ijk}L_k\,,
 \qquad
\bigl\{L_i\,,\g_j\,\bigr\}=\varepsilon_{ijk}\g_k \,,
\qquad
\bigl\{\g_i\,,\g_j\,\bigr\}=0\,.
\end{equation}
So, we can prove that the original nonholonomic  Chaplygin system is trajectory equivalent
to the dual integrable dynamical system on two-dimensional sphere $S^2$, which is a Hamiltonian system with respect to the
canonical Lie-Poisson brackets (\ref{e3}), see details in  \cite{ts12}.

\subsection{Generalized Chaplygin sphere, $\kappa=-1$\,.}
Now let us compare known Poisson structure at $\kappa=1$ with the new Poisson structure obtained for the case $\kappa=-1$. It is easy to see that at $\kappa=1$  bivector $P_g$ associated with the Poisson brackets (\ref{3-br}) has the form
\begin{equation}\label{p3}
 P_g=\mathrm g\,\left(\begin{array}{cc}0&\mathbf \Gamma\\ -\mathbf \Gamma^\top &
\mathbf M\end{array}\right)-d\mathrm g^{-1}\,(M,\mathbf A\g)\left(\begin{array}{cc}0&0\\ 0&
\mathbf \Gamma\end{array}\right)\,,
\end{equation}
where
\[
\mathbf \Gamma=\left(
                 \begin{array}{ccc}
                   0 & \g_3 & -\g_2 \\
                   -\g_3 & 0 & \g_1 \\
                   \g_2 & -\g_1 & 0
                 \end{array}
               \right)\,,\qquad
\mathbf M=\left(
                 \begin{array}{ccc}
                   0 & M_3 & -M_2 \\
                   -M_3 & 0 & M_1 \\
                   M_2 & -M_1 & 0
                 \end{array}
               \right)\,.
\]
Of course, canonical Poisson bivector on  $e^*(3)$
 \bq\label{can-p}
 P=\left(\begin{array}{cc}0&\mathbf \Gamma\\ -\mathbf \Gamma^\top &
\mathbf M\end{array}\right)\,,
 \eq
is compatible with its trivial deformation  $P_g$ so that
 \[[P,P_g]=0\,.\]
The Poisson bivector $P_b$ for the generalized Chaplygin ball rolling over the sphere,
albeit on the similar form, has completely another properties.
\begin{prop}
At $\kappa=-1$ integrals of motion  (\ref{3-int}-\ref{4-int}) are in  involution with
respect to the Poisson brackets defined by the following Poisson bivector
\begin{equation}\label{p3bmf}
 P_{b}=\mathrm g\,
 \left(\begin{array}{cc}0&\hat{\mathbf \Gamma}\\
 \\
                 -\hat{\mathbf \Gamma}^\top & \hat{ \mathbf M}\end{array}\right)
+\mathrm g^{-1}\,(2d(\g,\g)-\mathrm{tr}\mathbf B)\left(\begin{array}{cc}0&0\\ 0&
\widetilde{\mathbf \Gamma}\end{array}\right)\,,
\end{equation}
which is just one linear in $M$  solution of the equations (\ref{geom-eq}). Matrix $\hat {\mathbf \Gamma}$ depends only on $\g$
\[
\hat{\mathbf \Gamma}=\left((\g,\g)\,\mathbf E-\mathbf C-\dfrac{\mathbf B}{2d}\right)\mathbf \Gamma_b\,,\qquad \]
where $\mathbf E$ is a unit matrix,
\[
\mathbf C=
\left(
  \begin{array}{ccc}
    \g_1^2 & \g_1\g_2 & \g_1\g_3 \\
   \g_2\g_1 &\g_2^2 & \g_2\g_3 \\
  \g_3\g_1 & \g_3\g_2 & \g_3^2
  \end{array}
\right)
\quad\mbox{and}\qquad
\mathbf \Gamma_b=\left(
                 \begin{array}{ccc}
                   0 & b_3\g_3 & -b_2\g_2 \\
                   -b_3\g_3 & 0 & b_1\g_1 \\
                   b_2\g_2 & -b_1\g_1 & 0
                 \end{array}
               \right)\,.
\]
Entries of the other two matrices are equal to
\ben
\hat{\mathbf M}_{ij}&=&-\varepsilon_{ijk}\left(\alpha_k
\g_k-(\g,\g)b_kM_k+\dfrac{b_k^2M_k}{2d} \right)\,,\nonumber\\
\nonumber\\
\widetilde{\mathbf \Gamma}_{ij}&=&-\dfrac{\varepsilon_{ijk}b_k\g_k}{(b_1+b_2)(b_2+b_3)(b_1+b_3)}\Bigl(
(b_i+b_j)\,\alpha_k+(b_k-b_i)(b_k-b_j)M_k\g_k
\Bigr)\,,\nonumber
\en
here
\[
\alpha_k=\Bigl(C_2+b_k(\g,M)\Bigr)\,,\qquad C_2=(\g,\mathbf B M)\,.
\]
\end{prop}
The corresponding Poisson brackets look like
\[
\begin{array}{ll}
\{\g_i,\g_j\}_{b}=0\,,&\\
\{M_1,\g_1\}_{b}=\mathrm g (b_2-b_3)\g_1\g_2\g_3\,,\quad&\{M_1,\g_2\}_{b}=\mathrm g\g_3\left(
b_3(\g_1^2+\g_3^2)+b_2\g_2^2-\dfrac{b_2b_3}{2d}\right)\\
\\
&\{M_1,\g_3\}_{b}=-\mathrm g\g_2\left(
b_2(\g_1^2+\g_2^2)+b_3\g_3^2-\dfrac{b_2b_3}{2d}\right)\\
\\
\{M_2,\g_2\}_{b}=\mathrm g (b_3-b_1)\g_1\g_2\g_3\,,\quad&
\{M_2,\g_1\}_{b}=-\mathrm g\g_3\left(
b_1\g_1^2 +b_3(\g_2^2+\g_3^2)-\dfrac{b_1b_3}{2d}\right)\\
\\
&\{M_2,\g_3\}_{b}=\mathrm g \g_1\left(
b_1(\g_1^2+\g_2^2)+b_3\g_3^2-\dfrac{b_1b_3}{2d}
\right)\\
\\
\{M_3,\g_3\}_{b}=\mathrm g (b_1-b_2)\g_1\g_2\g_3\,,\quad&
\{M_3,\g_1\}_{b}=\mathrm g\g_2\left(
b_1\g_1^2+b_2(\g_2^2+\g_3^2)-\dfrac{b_1b_2}{2d}
\right) \\
\\
&\{M_3,\g_2\}_{b}=-\mathrm g\g_1\left(
b_(\g_1^2+\g_3^2)+b_2\g_2^2-\dfrac{b_1b_2}{2d}\right)\,,
\end{array}
\]
and
\[\begin{array}{lr}
\{M_1,M_2\}_b=&
\mathrm g\left(-\alpha_3\g_3+(\g,\g) b_3M_3-\frac{b_3^2M_3}{2d}\right)
-\mathrm g^{-1}\g_3\bigl(2d(\g,\g)-\mathrm{tr}\mathbf B\bigr)\times\qquad\qquad\\
\\
&\times
\left(\frac{\alpha_3b_3}{(b_3+b_2)(b_3+b_1)}+\frac{(b_3-b_2)(b_3-b_1)b_3\g_3M_3}{(b_3+b_2)(b_1+b_3)(b_2+b_1)}\right)\,.\nonumber
\end{array}
\]
Brackets  $\{M_1,M_3\}_b$ and  $\{M_2,M_3\}_b$ have the same form as  $\{M_1,M_2\}_b$ and, therefore,
we omit their explicit expressions.

If $C_2=0$ there are many other linear in momenta $M$ solutions of  the equations (\ref{geom-eq}) associated with known variables of separation, see details in \cite{ts12rd}.

At $C_2\neq 0$ using  linear in momenta Poisson brackets $\{.,.\}_b$ we can rewrite equations of motion   (\ref{m-eq})
in the following form
 \begin{equation}\label{new-eq2}
\dfrac{\mathrm dx_k}{\mathrm dt}\equiv X_k=\mathrm g_1^{-1}\{H_1,x_k\}_b+\mathrm g_2^{-1}\{H_2,x_k\}_b\,,
\end{equation}
where
 \[
\mathrm g_1(\g)=\dfrac{\mathrm g(\g)\,s(\g)}{(2d(\g,\g)-\mbox{\rm tr}\mathbf B)d}\,,\qquad
\mathrm g_2(\g)=\dfrac{\mathrm g(\g)\,s(\g)}{2d}\]
and
\bq\label{s-fun}
s(\g)= 4d^2(\g,\g)(\g,\mathbf B\g)-2d\bigl( (\mathbf E\,\mbox{tr}\mathbf B-\mathbf B)\g,\mathbf B\g\bigr)+\mbox{det}\mathbf B\,.
 \eq
It is easy to see that at $\kappa=-1$ equations of motion have a more complicated from
in comparison with  the original Chaplygin problem (\ref{new-eq}) at $\kappa=1$.

 \begin{prop}
At $\kappa=-1$  the initial non-Hamiltonian vector field $X$ (\ref{m-eq},\ref{new-eq2}) is a sum of
two conformally Hamiltonian vector fields
\[X =\mathrm g_1^{-1}\,\hat{X}_1+\mathrm g_2^{-1}\,\hat{X}_2\,,\]
where $\hat{X}_{1,2}$ are hamiltonian vector fields associated with two commuting integrals of motion
\[
 \hat{X}_{1}=P_b\,\mathrm dH_{1}\,, \qquad \hat{X}_{2}=P_b\,\mathrm dH_{2}\,,\qquad \{H_1,H_2\}_b=0\,.
\]
On the other hand,  this non-Hamiltonian vector field is conformally Hamiltonian vector field
\bq\label{conf-bfm}
X={\mathrm g}_3^{-1} \hat{X}_3\,,\qquad\mbox{where}\qquad \hat{X}_3=P_b\mathrm d H_3\qquad\mbox{and}\qquad \mathrm g_3(\g)=\dfrac{\mathrm g(\g) s(\g)}{d}\,,
\eq
but with respect to another Hamiltonian
\bq\label{h-bfm}
H_3=(2d(\g,\g)-\mbox{\rm tr}\mathbf B)H_1+2H_2\,,
\eq
which is an integral of motion of  (\ref{m-eq}) without
any distinguished physical meaning.
\end{prop}
At  $\kappa=1$ the similar linear combination
$\hat{H}_3=H_2-dH_1$ coincides with the Hamiltonian of the Veselova system, which is equivalent to
the original Chaplygin ball at the special choice of parameters $a_i$ \cite{ts12}.

Now let us discuss the main difference between bivectors $P_g$ and $P_b$.
\begin{prop}
Bivector $P_b$ (\ref{p3bmf}) is nontrivial deformation of the canonical Poisson bivector $P$ (\ref{can-p}) on the Lie algebra $e^*(3)$.
\end{prop}
In contrast with bivector $P_g$ (\ref{p3}) bivector  $P_b$ (\ref{p3bmf}) is  incompatible with
the ca\-no\-ni\-cal Poisson bi\-vec\-tor $P$ (\ref{can-p}) on $e^*(3)$ because
\[
[P,P_g]=0\,,\qquad\mbox{whereas}\qquad [P,P_b]\neq 0\,.
\]
As sequence bivector $P_b$ can not be trivial deformation of $P$.

Nevertheless, bivector $P_b$ is nontrivial deformation of canonical bivector $P$, because there is change of variables
\ben
L_1&=&\dfrac{1}{\mathrm g(\g)s(\g)b_1b_2(\g_1^2+\g_2^2)}\left(\alpha_1(b_{1}\g_2M_1-b_2\g_1M_2)+\beta_1M_3+\dfrac{b\g_1\g_3h(\g)}{\g_1^2+\g_2^2}\right)+\dfrac{c\g_1}{\g_1^2+\g_2^2}\,,
\nonumber\\
L_2&=&\dfrac{1}{\mathrm g(\g)s(\g)b_1b_2(\g_1^2+\g_2^2)}\left(\alpha_2(b_{1}\g_2M_1-b_2\g_1M_2)+\beta_2M_3+\dfrac{b\g_2\g_3h(\g)}{\g_1^2+\g_2^2}\right)+\dfrac{c\g_2}{\g_1^2+\g_2^2}\,,\nonumber\\
\nonumber\\
L_3&=&\dfrac{1}{\mathrm g(\g)s(\g)b_1b_2(\g_1^2+\g_2^2)}\Bigl(\alpha_3(b_{1}\g_2M_1-b_2\g_1M_2)+\beta_3M_3-bh(\g)\Bigr)\,,\nonumber\\
\nonumber\\
b&=&(B\g,M)\,,\qquad c=(\g,L)\,,\nonumber
\en
which allows us to  reduces this bivector to canonical one. Here $s(\g)$ is given by (\ref{s-fun}),
\ben
\alpha_1&=&2d\g_2\bigl(2db_1\g_1^2+2db_2(\g_2^2+\g_3^2)-b_1b_2\bigr)\,,\qquad
\alpha_3=4d^2\g_1\g_2\g_3(b_1-b_2)\,,\nonumber\\
\nonumber\\
\alpha_2&=&-2d\g_1\bigl(2db_1(\g_1^2+\g_3^2)+2db_2\g_2^2-b_1b_2\bigr)\,,\nonumber
\en
and
\ben
\beta_1&=&-2d\g_1\g_3\Bigl(2d\bigl(\g_1^2(b_1b_2-b_1b_3+b_2b_3)+b_2(b_1\g_1^2+b_3\g_3^2)\bigr)-b_1b_2b_3\Bigr)\,,\nonumber\\
\nonumber\\
\beta_2&=&-2d\g_2\g_3\Bigl(2d\bigl(\g_1^2(b_1b_3+b_1b_2-b_2b_3)+b_1 (b_2\g_2^2+b_3\g_3^2)\bigr)-b_1b_2b_3\Bigr)\,,\nonumber\\
\nonumber\\
\beta_3&=&4d^2b_3\g_3^2(b_2\g_1^2+b_1\g_2^2)+2db_1b_2(\g_1^2+\g_2^2)\bigl(2d(\g_1^2+\g_2^2)-b_3\bigr)
\,.\nonumber
\en
For the brevity we omit the explicit expressions for the function
 $h(\g)$, which is a solution of the differential equations $\bigl\{L_i\,,L_j\,\bigr\}_b=\varepsilon_{ijk}L_k\,$.

Applying this transformation at $c=(\g,L)=0$ to the Hamilton function (\ref{h-bfm}) one gets  integrable dynamical system on the two-dimensional sphere, which is a standard Hamiltonian system with respect to canonical Poisson brackets on $T^*S^2$. The generalised nonholonomic  Chaplygin sphere is trajectory equivalent to this Hamiltonian system up to change of time defined by (\ref{conf-bfm}).

\end{document}